\documentclass[twocolumn,aps,showpacs,superscriptaddress,pre]{revtex4}

\usepackage{amsmath,amssymb,graphicx}
\usepackage{bm}
\usepackage{stmaryrd}
\usepackage{times}
\usepackage{color}

\newcommand{\W}{W} 
\newcommand{\pbc}{\pi}
\newcommand{\abc}{\overline{\pi}}
\newcommand{\lav}{\bm{\langle}}
\newcommand{\rav}{\bm{\rangle}}
\newcommand{\cJ}{{\cal J}}
\newcommand{\GD}{\Gamma_{\ell}}
\newcommand{\PJ}{P_{\cal J}(q)}
\newcommand{\PD}{P(q)}

\newcommand{\ds}{d_{\rm s}}
\newcommand{\GS}{\Gamma_{\rm s}}

\definecolor{yblue}{rgb}{0.06, 0.3, 0.57}
\usepackage[pdftex]{hyperref}
\hypersetup{colorlinks=true,linkcolor=blue,citecolor=blue,urlcolor=blue}

\begin{document}

\title{Number of thermodynamic states in the three-dimensional
Edwards-Anderson spin glass}

\author{Wenlong Wang}
\affiliation{Department of Physics and Astronomy, Texas A$\&M$ University, 
College Station, Texas 77843-4242, USA}

\author{Jonathan Machta}
\email{machta@physics.umass.edu}
\affiliation{Department of Physics, University of Massachusetts,
Amherst, Massachusetts 01003 USA}
\affiliation{Santa Fe Institute, 1399 Hyde Park Road, Santa Fe, New Mexico
87501, USA}

\author{Humberto Munoz-Bauza}
\affiliation{Department of Physics and Astronomy, Texas A$\&M$ University, 
College Station, Texas 77843-4242, USA}

\author{Helmut G. Katzgraber}
\affiliation{Department of Physics and Astronomy, Texas A$\&M$ University, 
College Station, Texas 77843-4242, USA}
\affiliation{Santa Fe Institute, 1399 Hyde Park Road, Santa Fe, New Mexico 
87501, USA}

\begin{abstract}

The question of the number of thermodynamic states present in the
low-temperature phase of the three-dimensional Edwards-Anderson Ising
spin glass is addressed by studying spin and link overlap distributions
using population annealing Monte Carlo simulations. We consider overlaps
between systems with the same boundary condition---which are the usual
quantities measured---and also overlaps between systems with different
boundary conditions, both for the full systems and also within a smaller
window within the system. Our results appear to be fully compatible with
a single pair of pure states such as in the droplet/scaling picture.
However, our results for whether or not domain walls induced by changing
boundary conditions are space filling or not are also compatible with
scenarios having many thermodynamic states, such as the chaotic pairs
picture and the replica symmetry breaking picture. The differing results
for spin overlaps in same and different boundary conditions suggest that
finite-size effects are very large for the system sizes currently
accessible in low-temperature simulations.

\end{abstract}

\maketitle

\section{Introduction}

The nature of the low-temperature phase of Ising spin glasses is a
long-standing mystery and the subject of considerable controversy
\cite{mezard:84,moore:98,drossel:00,krzakala:00,palassini:00,marinari:00,marinari:00a,marinari:00c,katzgraber:01,middleton:01,hatano:02,katzgraber:02,katzgraber:03,katzgraber:03f,hed:07,leuzzi:08,alvarez:10a,yucesoy:12,billoire:13,yucesoy:13b,wang:14}.
A central question is whether the low-temperature phase is comprised of
a single pair of pure thermodynamic states, or whether the situation is
more complicated and involves an infinite number of pure states
\cite{parisi:79}. For many years, efforts to characterize the
low-temperature phase were hampered by confusion over the concept of the
thermodynamic limit for systems---such as spin glasses---that may
display chaotic size dependence. Chaotic size dependence means that
different thermodynamic states may appear in an observation volume in
different system sizes or with different boundary conditions. If chaotic
size dependence occurs, the usual thermodynamic limit is not meaningful
and must be replaced by the so-called {\em metastate}
\cite{aizenman:90,newman:96a,newman:97}, which is a probability
distribution over thermodynamic states that may be observed in the
observation volumes as the system size changes. Newman and Stein showed
that the properties of the metastate are severely constrained for
finite-dimensional spin glasses. The metastate may either be trivial and
contain a single thermodynamic state consisting of a pair of pure states
related by a global spin flip or the metastate may have support on a
uncountable infinity of thermodynamic states. The {\em droplet picture},
developed by McMillan \cite{mcmillan:84b}, Bray and
Moore~\cite{bray:86}, as well as Fisher and
Huse~\cite{fisher:86,fisher:87,fisher:88}, is an example of the simple
scenario where the metastate consists of a single pair of pure states
and the thermodynamic limit may be defined in the usual way. There are
two plausible ways that a metastate with support on a continuum of
thermodynamic states could occur: Either the {\em chaotic
pairs}~\cite{newman:92,newman:97} scenario where in every finite volume
only a single pair of pure states is manifest, or the {\em nonstandard
replica symmetry breaking} (RSB) scenario \cite{newman:97,read:14a},
where in every volume a countably infinite number of thermodynamic pure
states are manifest. The nonstandard RSB scenario is the
finite-dimensional analog of Parisi's {\rm replica symmetry breaking
solution} \cite{parisi:79,parisi:80,parisi:83} of the mean-field
Sherrington-Kirkpatrick model \cite{sherrington:75}, i.e., the Ising
spin glass on the complete graph. The idea that behavior similar to
Parisi's RSB solution of the Sherrington-Kirkpatrick model also applies
to the three-dimensional Ising spin glass has a long
history~\cite{parisi:79,parisi:80,parisi:83,rammal:86,mezard:87,young:98,parisi:08};
the nonstandard RSB scenario being the only mathematically well-defined
realization of the mean-field RSB theory in finite dimensions.

One of the key differences between scenarios with many thermodynamic
states and a single pair of pure states is whether domain walls induced
by changing boundary conditions are space filling. If there is a single
thermodynamic state composed of a pair of pure states related by a
global spin flip, then the relative domain wall induced by changing from
periodic to anti-periodic boundary conditions in a single direction is
not space filling. Given an observation window within a much larger
system, this domain wall deflects out of the observation window and a
single pure state will be seen in the observation volume as the larger
volume is taken to infinity. On the other hand, if there is chaotic size
dependence and many thermodynamic states, then a change in boundary
conditions is likely to induce a completely different thermodynamic
state and the relative domain wall will be space filling.

Many numerical studies have attempted to discern which theoretical
picture correctly describes short-range systems (see, for example,
Refs.~\cite{mezard:84,moore:98,drossel:00,krzakala:00,palassini:00,marinari:00,marinari:00a,marinari:00c,katzgraber:01,middleton:01,hatano:02,katzgraber:02,katzgraber:03,katzgraber:03f,hed:07,leuzzi:08,alvarez:10a,yucesoy:12,billoire:13,yucesoy:13b,wang:14})
resulting in often contradictory conclusions. Most notably, studies of
the average distribution of the spin overlap---the order parameter for
spin glasses---find a finite weight near zero overlap
\cite{palassini:00,marinari:00,katzgraber:01,katzgraber:02}, suggesting
a large multiplicity of pure states. However, because finite-size
corrections are expected to be severe for currently-available system
sizes in simulations, new methods that go beyond simple disorder
averaging have been developed to distinguish the different competing
pictures
\cite{yucesoy:12,middleton:13,monthus:13,wang:14,wittmann:14a}.
These analyses, in contrast, suggest a thermodynamic limit with only a
single pair of pure states.

To further investigate this problem, in this work we set out to
determine whether relative domain walls induced by changing boundary
conditions (i.e., overlaps between configurations with different
boundary conditions) are space filling and, if not, we measure their
fractal dimension directly. To this end, we carry out large-scale Monte
Carlo simulations to study the spin and link overlap distributions
\cite{katzgraber:01} and measure these overlaps both in a single
boundary condition and between spin configurations chosen from different
boundary conditions. These measurements are made for both the full
volume of the studied system, as well as for an observation window
smaller than the full volume. Similar ideas have also been employed in
previous work. For example, spin overlap distributions in a small window
with the same boundary condition were known to be similar to those in
the full system, which has been taken as a support of RSB in
Ref.~\cite{marinari:98f}. But when spin overlap functions in a small
window between different boundary conditions are studied, this is no
longer clear. The spin overlap and link overlap functions between
different boundary conditions at zero temperature have also been used in
determining whether domain walls are space filling in
Refs.~\cite{palassini:99,marinari:00}, however these two studies reach
opposite conclusions. Here we revisit this problem systematically at
nonzero temperatures. Our results for the spin overlap from differing
boundary conditions point toward scenarios having a single thermodynamic
state consisting of a pair of pure states (as is the case for the
droplet/scaling picture). However, our results for the domain wall
induced by changing boundary conditions are also compatible with many
pairs of pure states as is the case for the RSB and chaotic pairs
picture. We therefore conclude that either accessible system sizes in
simulations are still too small despite novel analysis techniques or
the nature of the spin-glass state is more complex than expected.

The paper is organized as follows. We first discuss the model,
observables and simulation methods in Sec.~\ref{model}, followed by
numerical results in Sec.~\ref{results}. Concluding remarks are stated
in Sec.~\ref{summary}.

\section{Model, observables and methods}
\label{model}

We study the three-dimensional Edwards-Anderson (EA) Ising spin-glass model
\cite{edwards:75} defined by the Hamiltonian
\begin{equation}
H = - \sum_{\langle ij \rangle} J_{ij} S_i S_j ,
\label{eq:ham}
\end{equation}
where $S_i=\pm 1$ are Ising spins and the sum is over nearest neighbors
on a cubic lattice of linear size $L$. The random couplings $J_{ij}$
are chosen from a Gaussian distribution with mean $0$ and variance $1$.
A set of couplings $\cJ =\{ J_{ij} \}$ defines a disorder realization.

We study two primary observables, the spin overlap $q$ and the link
overlap $q_\ell$ defined, respectively, as
\begin{equation}
q=\dfrac{1}{\lambda^3}\sum\limits_i S_{i}^{(1)} S_{i}^{(2)},
\label{e2}
\end{equation}
and
\begin{equation}
{q_\ell}=\dfrac{1}{dL^3}\sum\limits_{\langle ij \rangle} 
S^{(1)}_{i} S^{(1)}_{j} S^{(2)}_{i} S^{(2)}_{j},
\label{e3}
\end{equation}
where spin configurations ``(1)'' and ``(2)'' are chosen independently
from the Boltzmann distribution. The sum in Eq.~\eqref{e2} is either
over the full lattice with $L \times L \times L$ sites, or a smaller
observation window of size $\W \times \W \times \W$ with $\W < L$, and,
for these two cases, $\lambda=L$ and $\lambda=W$, respectively. The link
overlap is measured only for the full lattice and the sum in
Eq.~\eqref{e3} is over nearest neighbors. $d = 3$ is the space
dimension.  Two boundary conditions are studied: (i) periodic boundary
conditions, referred to as $\pbc$, and (ii) periodic boundary conditions
in the $y$ and $z$ directions with anti-periodic boundary conditions in
the $x$ direction, referred to as $\abc$. The two boundary conditions
differ in whether $S_{\vec{r}+\hat{x}L}= \pm S_{\vec{r}}$ with the $+$
sign for $\pbc$ and the $-$ sign for $\abc$. Averages over the Gibbs
distribution are indicated with angular brackets and the boundary
condition for the two copies required in the definition of an overlap
are specified by subscripts. For example, $\langle q_\ell
\rangle_{\pbc,\abc}$ is the link overlap between configurations $(1)$
and $(2)$ in boundary conditions $\pbc$ and $\abc$, respectively. We
refer to overlaps between the same boundary condition, either
$(\pbc,\pbc)$ or $(\abc,\abc)$, as {\em diagonal} overlaps and overlaps
between different boundary conditions, $(\pbc,\abc)$ or $(\pbc,\abc)$,
as {\em off-diagonal} overlaps.

We use population annealing Monte Carlo (PAMC)
\cite{hukushima:03,zhou:10,machta:10,wang:15e} to carry out the
simulations. In PAMC, a large population of $R_0$ replicas of the
system, each with the same disorder realization, are annealed in
parallel from infinite temperature to a low target temperature, $T_0 =
1/\beta_0$. The annealing schedule consists of a sequence of $N_T$
temperatures equally spaced in inverse temperature $\beta$. In a
temperature step from $\beta$ to $\beta^\prime$ the population is
resampled; some members of the population are eliminated and some are
reproduced. The mean number of copies of replica $i$ is proportional to
the re-weighting factor, $\exp[-(\beta^\prime-\beta) E_i]$. The constant
of proportionality is chosen so that the population size remains close
to $R_0$. Following each resampling step, the population at $\beta'$ is
acted on by $N_S = 10$ lattice sweeps of the Metropolis algorithm. We
simulate $N_{\rm sa}$ disorder realizations and measure overlaps at $T =
T_0 = 0.2$ and $T=0.42$. Both of these temperatures are deep in the
low-temperature spin-glass phase and should therefore not be affected by
critical fluctuations.  Overlaps are then measured by pairing
independent replicas in the population. The simulation parameters are
listed in Table~\ref{table}.  A small number of disorder realizations
that were not equilibrated using population size listed in Table
\ref{table} were re-run with larger populations until equilibration was
achieved.

\begin{table}
\caption{
Parameters of the simulations for linear system sizes $L$ and for both
periodic ($\pbc$) and anti-periodic ($\abc$) boundary conditions. $R_0$
is the population size, $T_0 = 1/\beta_0$ is the lowest temperature
simulated, $N_T$ the number of temperatures used in the annealing
schedule, and $N_{\rm sa}$ the number of disorder realizations.
\label{table}
}
\begin{tabular*}{\columnwidth}{@{\extracolsep{\fill}} l c c c r}
\hline
\hline
$L$  & $R_0$ & $1/\beta_0$ & $N_T$  & $M$ \\
\hline
$4$  & $5\,10^4$ & $0.200$     & $101$   & $4891$ \\
$6$  & $2\,10^5$ & $0.200$     & $101$   & $5000$ \\
$8$  & $5\,10^5$ & $0.200$     & $201$   & $4844$ \\
$10$ & $   10^6$ & $0.200$     & $301$   & $4600$ \\
$12$ & $   10^6$ & $0.333$     & $301$   & $4137$ \\
\hline
\hline
\end{tabular*}
\end{table}

\section{Results}
\label{results}

We discuss the spin overlap in Sec.~\ref{spinoverlap} and the link
overlap in Sec.~\ref{linkoverlap}. When possible, we carry out fits of
the data assuming either space filling and non-space filling domain
walls and compare the quality of the fits.

\subsection{Spin overlap}
\label{spinoverlap}

\begin{figure}[htb]
\begin{center}
\includegraphics[width=\columnwidth]{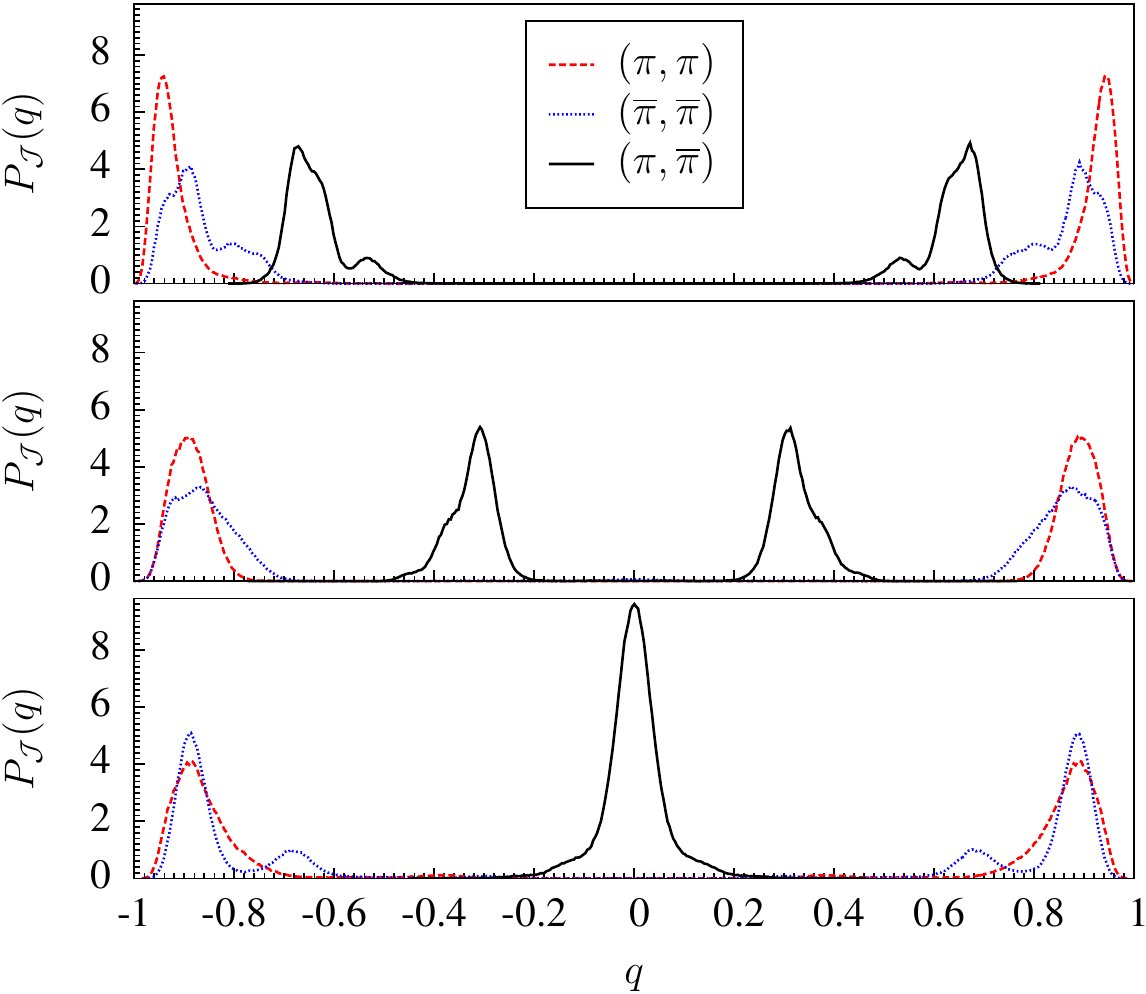}
\caption{
Spin overlap distributions $\PJ$ for three typical disorder realizations
for $L = 8$ at $T = 0.42$.  Each panel represents a different disorder
realization and, within each panel, the overlap is shown for the
boundary condition pairs $(\pbc ,\pbc)$ (red, dashed),  $(\abc ,\abc)$
(blue, dotted) and $(\pbc, \abc)$ (black, solid). The peaks in $\PJ$ for
the off-diagonal pair are shifted away from $q_{\rm EA}$ due to the
relative domain wall between the periodic and anti-periodic boundary
conditions.
}
\label{Pq3}
\end{center}
\end{figure}

The spin overlap distributions $\PJ$ for three typical disorder
realizations for the full system ($L=8$ and $T=0.42$) are shown in each
of the three panels of Fig.~\ref{Pq3}. In each panel, the three curves
represent the three boundary condition pairs $(\pbc ,\pbc)$, $(\abc
,\abc)$, and $(\pbc, \abc)$.  Observe that the two diagonal spin
overlaps, $(\pbc ,\pbc)$ and $(\abc ,\abc)$, each have peak(s) near the
finite-size values of the Edwards-Anderson order parameter, $\pm q_{\rm
EA}$. On the other hand, the off-diagonal overlap distributions have
peaks that are shifted closer to the origin (i.e., $|q| < q_{\rm
EA}$) because the domain wall induced by changing boundary conditions
reduces the probability of a large value of the spin overlap.

\begin{figure}[htb]
\begin{center}
\includegraphics[width=\columnwidth]{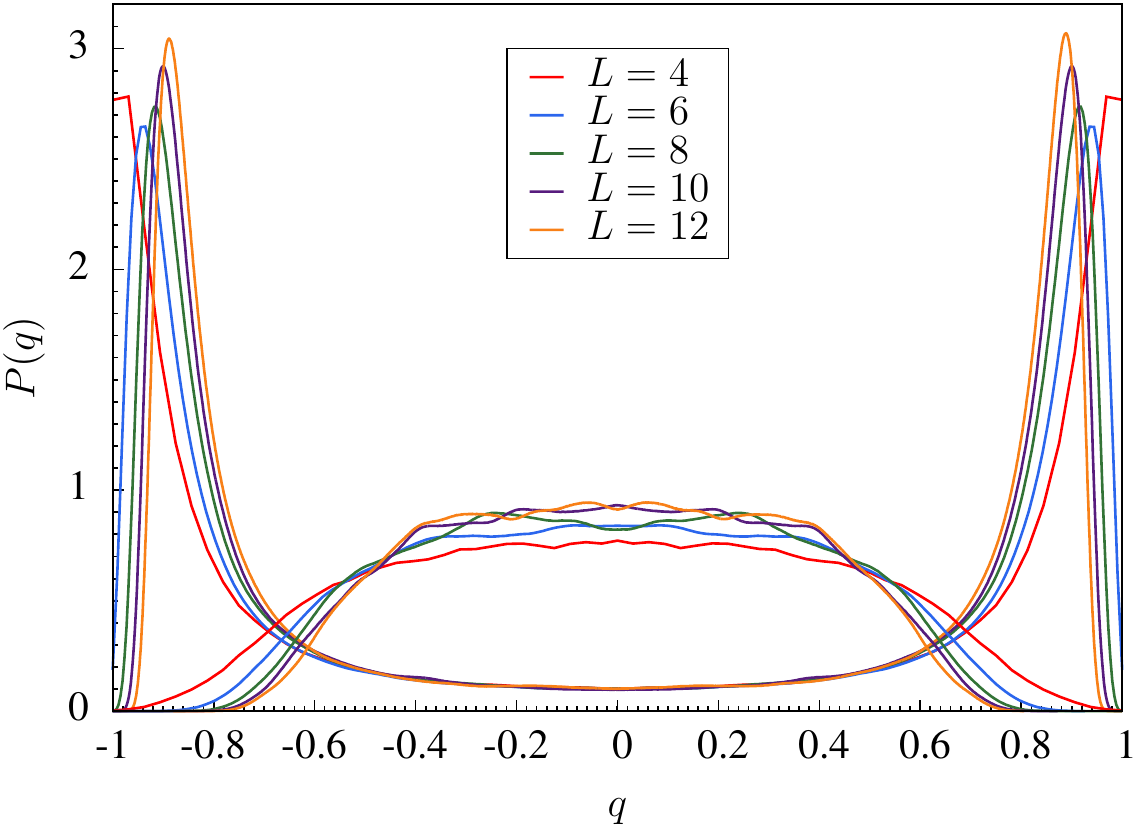}
\caption{
Disorder-averaged spin overlap distributions $P(q)$ for the full lattice
for sizes $L=4$, $6$, $8$, $10$, and $12$ at $T=0.42$. The set of curves
with peaks at the finite-size value of $\pm q_{\rm{EA}}$ and bimodal
features correspond to the diagonal overlap distributions while the set
of curves with broad maxima at the center correspond to the off-diagonal
overlap distributions.
}
\label{PqN}
\end{center}
\end{figure}

Figure \ref{PqN} shows $\PD$, the disorder averaged spin overlap
distribution for sizes $L=4$, $6$, $8$, $10$, and $12$ at $T=0.42$ for
the full lattice. The diagonal overlap distribution displays peaks at
finite-size values of $\pm q_{\rm EA}$ with $q_{\rm EA}$ decreasing with
$L$ while for small $q$ the distribution is nearly independent of $L$,
consistent with past studies
\cite{palassini:00,katzgraber:01,katzgraber:02,yucesoy:12}. The
off-diagonal overlap distribution, however, has {\em no} peaks close to
$\pm q_{\rm EA}$. Instead, the curves are nearly independent of $L$ near
$q=0$ and seem to approach a relatively flat distribution bounded by
$\pm q_{\rm{EA}}$. It is instructive to compare this with the
ferromagnetic Ising model where the domain wall dimension is $d-1$.
Because a domain-wall can be inserted anywhere in the system, $P(q)$
would have a flat distribution between for $|q| \leq q_{EA}$. If a
single thermodynamic state picture such as droplet scaling is correct
for the EA model, the domain wall is expected to be fractal and the
off-diagonal $P(q)$ has a broad maximum at $q=0$ with $P(q)$ decreasing
toward $\pm q_{EA}$ just as is seen in Fig.~\ref{PqN}.  On the other
hand, for many state scenarios such as RSB, changing boundary conditions
almost always results in a completely different thermodynamic state so
that the off-diagonal overlap distribution displays a $\delta$-function
at the origin in the infinite-volume limit \cite{aspelmeier:16a}. We see
no evidence of a $\delta$-function at $q=0$, and the behavior of the
off-diagonal spin overlap therefore suggests a single pair of pure
states.

\begin{figure}[htb]
\begin{center}
\includegraphics[width=\columnwidth]{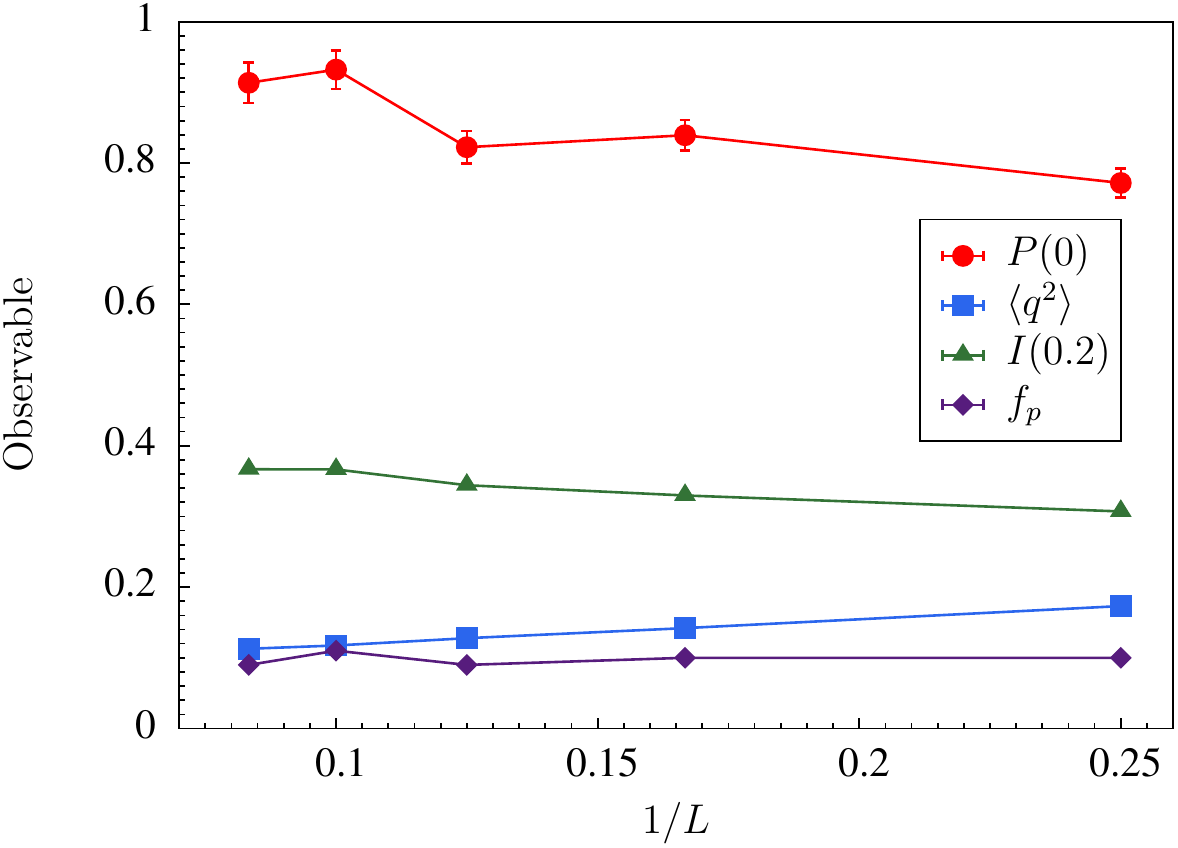}
\caption{
Four statistics of the off-diagonal overlap distribution: $P(0)$,
$\langle q^2 \rangle$, $I(0.2)$, and $f_q$ (see text for details) vs
inverse system size $1/L$. The expected thermodynamic behavior, i.e.,
$1/L \to 0$ is described in Eqs.~\eqref{eq:filling} and
\eqref{eq:fractal}.
}
\label{Statistics}
\end{center}
\end{figure}

To probe the behavior of the off-diagonal spin overlap distribution more
quantitatively, we also analyze four statistics of $P(q)$:
\begin{enumerate}
\item[$\bullet$] $P(0)$, the value of $P(q)$ at $q=0$.
\item[$\bullet$] $\langle q^2 \rangle$, the second moment of $P(q)$.
\item[$\bullet$] $I(0.2)=\int_{|q| \le 0.2} P(q) dq$.
\item[$\bullet$] $f_q$, the fraction of disorder realizations having a peak in the overlap distribution centered at $q=0$.
\end{enumerate}
If the relative domain wall is {\em space filling}, we expect that in the
thermodynamic limit
\begin{eqnarray}
P(0) &\rightarrow& \infty, \nonumber \\ 
\langle q^2 \rangle &\rightarrow& 0, \nonumber \\
I(0.2) &\rightarrow& 1, \nonumber \\
f_q &\rightarrow& 1.
\label{eq:filling}
\end{eqnarray}
However, if the relative  domain wall is a {\em fractal}
\begin{eqnarray}
P(0) &<& \infty, \nonumber \\ 
\langle q^2 \rangle &>& 0, \nonumber \\
I(0.2) &<& 1, \nonumber \\
f_q &<& 1. 
\label{eq:fractal}
\end{eqnarray}
Note that disorder averages are implied in the previous equations.
The four statistics are shown in Fig.~\ref{Statistics}. One can see that
indeed our data for the off-diagonal overlap distributions are
compatible with a fractal relative domain wall and a single pair of pure
states, but are not compatible with multiple pure states and space
filling domain walls. The fact that the off-diagonal spin overlap
suggests a single thermodynamic state scenario while the diagonal spin
overlap suggests a thermodynamic scenario with multiple states implies
very large finite-size corrections. This means that the average spin
overlap cannot be used in currently accessible simulation system
sizes to distinguish between these fundamentally different scenarios.

\begin{figure}[htb]
\begin{center}
\includegraphics[width=\columnwidth]{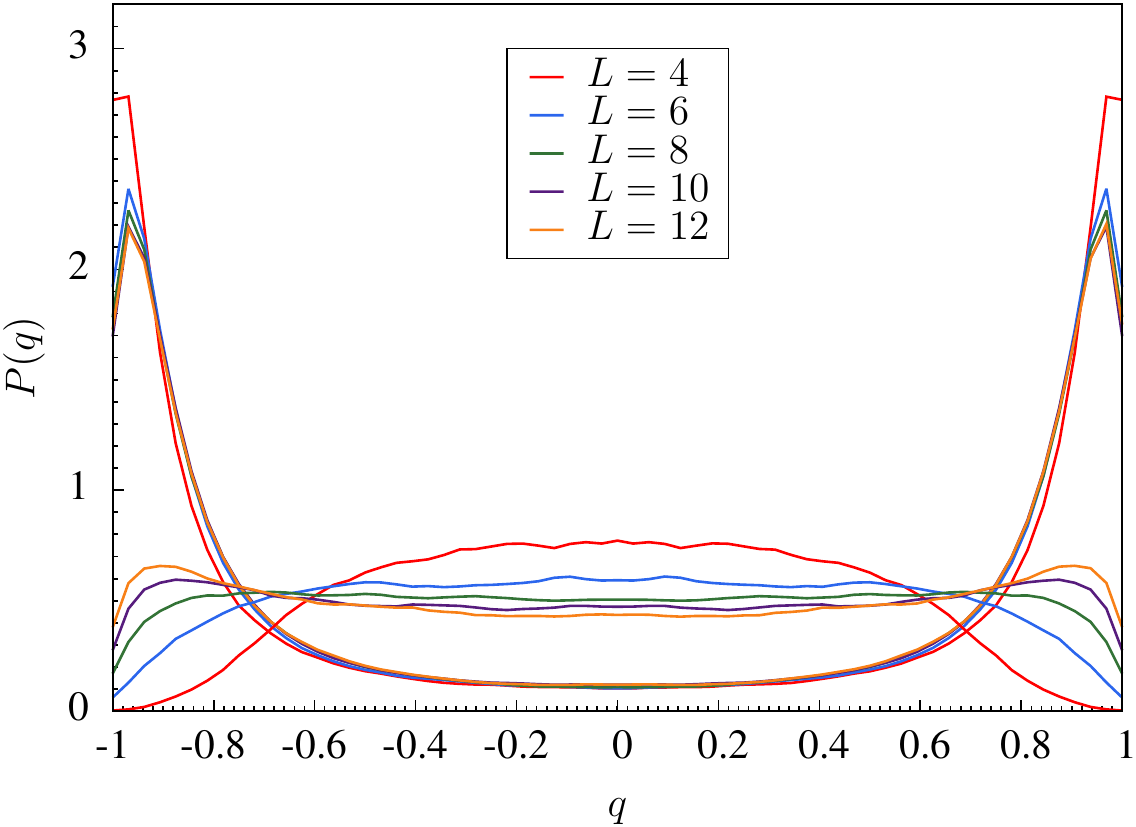}
\caption{
Disorder-averaged spin overlap distributions $P(q)$ in a $\W=4$
observation window within a system of size $L=4$, $6$, $8$, $10$, and
$12$ at $T=0.42$. The sets of curves that display large peaks near $\pm
q_{\rm EA}$, are the diagonal overlaps. The off-diagonal overlap curves
have much smaller weight near $\pm q_{\rm EA}$, but that weight
increases slowly with the system size $L$.
}
\label{Pq1}
\end{center}
\end{figure}

We next turn to the behavior of the spin overlap in an observation
window of size $\W$ within a system of size $L \geq W$.
Figure~\ref{Pq1} shows spin overlap distributions measured in a window
of size $\W=4$ within a system of size $L$ with $L=4$, $6$, $8$, $10$,
and $12$ at $T=0.42$. The curves with peaks at $\pm q_{\rm{EA}}$ are the
diagonal overlaps. These curves are only slightly dependent on the size
of the full system as first noted in Ref.~\cite{marinari:98f}.  On the
other hand, the off-diagonal spin overlaps do evolve significantly with
the size of the full system. It is notable that as $L$ increases, peaks
emerge at the finite-size value of the Edwards-Anderson order parameter.
This phenomenon was observed qualitatively for ground states in three
dimensions in Ref.~\cite{palassini:99} and suggests that the domain wall
induced by switching boundary conditions might not be space filling and
appears to deflect out of the window. If, as $L$ becomes large, the
off-diagonal and diagonal spin overlap become equal, it would be strong
evidence in favor of the droplet/scaling picture. However, much larger
system sizes would be needed to see such behavior.

To examine the behavior of the window overlap more quantitatively, we
compare integrals of the diagonal and off-diagonal spin overlap peaks in
the region near $\pm q_{\rm{EA}}$. We define $\GS$ as the difference
of the disorder averaged diagonal and off-diagonal overlap distributions
measured in the observation window and integrated over a set ${\cal A}$
\begin{equation}
\GS=\frac{1}{2}\!\int_{\cal A} dq 
\left[ \PD_{\pbc,\pbc} \!+\!\PD_{\abc,\abc} \!-\!
\PD_{\pbc,\abc} \!-\! \PD_{\abc,\pbc} \right]\! .
\end{equation}
We set ${\cal A}= [-1,-q_0) \cup (-q_0,1]$, where $q_0$ is chosen to
include the $\pm q_{\rm{EA}}$ peaks. A similar quantity was studied for
ground states in Ref.~\cite{palassini:99}.  If there is indeed a single
thermodynamic state and the domain wall deflects out of the window, then
$\GS$ should approach zero as $L \rightarrow \infty$. Furthermore, if
the domain wall is a fractal with fractal dimension $\ds$, then we
expect from box counting that 
\begin{equation}
\GS \sim (L/W)^{\ds-d} \to 0 \,\,\, {\rm for} \,\,\, L \rightarrow \infty , 
\label{eq:fitfrac}
\end{equation}
where $W$ is the window size, $\ds$ the fractal dimension of the domain
wall, and $d=3$ the spatial dimension. A natural cut-off for measuring
$\GS$ is to set $q_0 = q_c$, where $q_c$ is the crossing point of the
off-diagonal spin overlap seen in Fig.~\ref{Pq1}. For the $W=4$ window,
$q_c \approx 0.67$ for $T=0.42$ and $q_c \approx 0.69$ for $T=0.2$.
Note, however, that our results are insensitive to the choice of $q_0$.

\begin{figure}[htb]
\begin{center}
\includegraphics[width=\columnwidth]{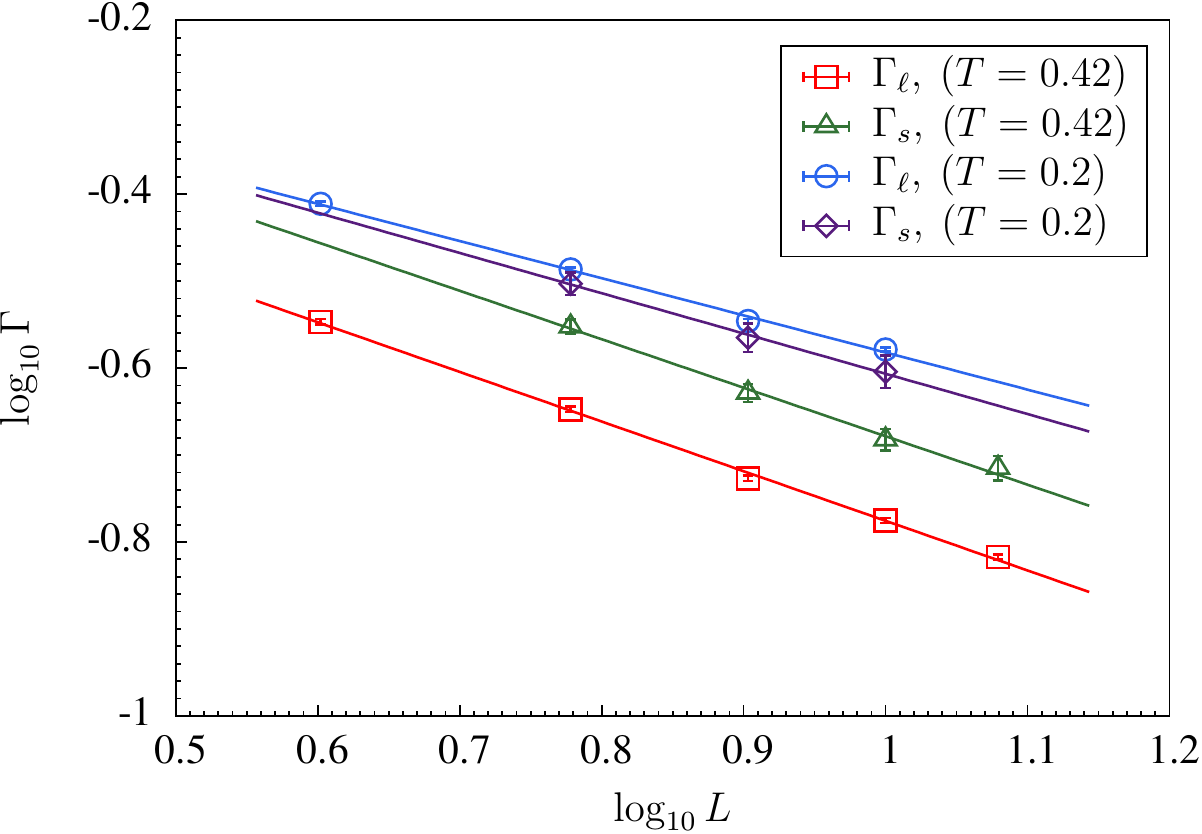}
\caption{
Scaling of $\GS$ (within a window) and $\GD$ (defined below, for the
full systems) as a function of system size $L$ for $T = 0.20$ and $T =
0.42$ together with power law fits (straight lines).
}
\label{Delta}
\end{center}
\end{figure}

\begin{figure}[htb]
\begin{center}
\includegraphics[width=\columnwidth]{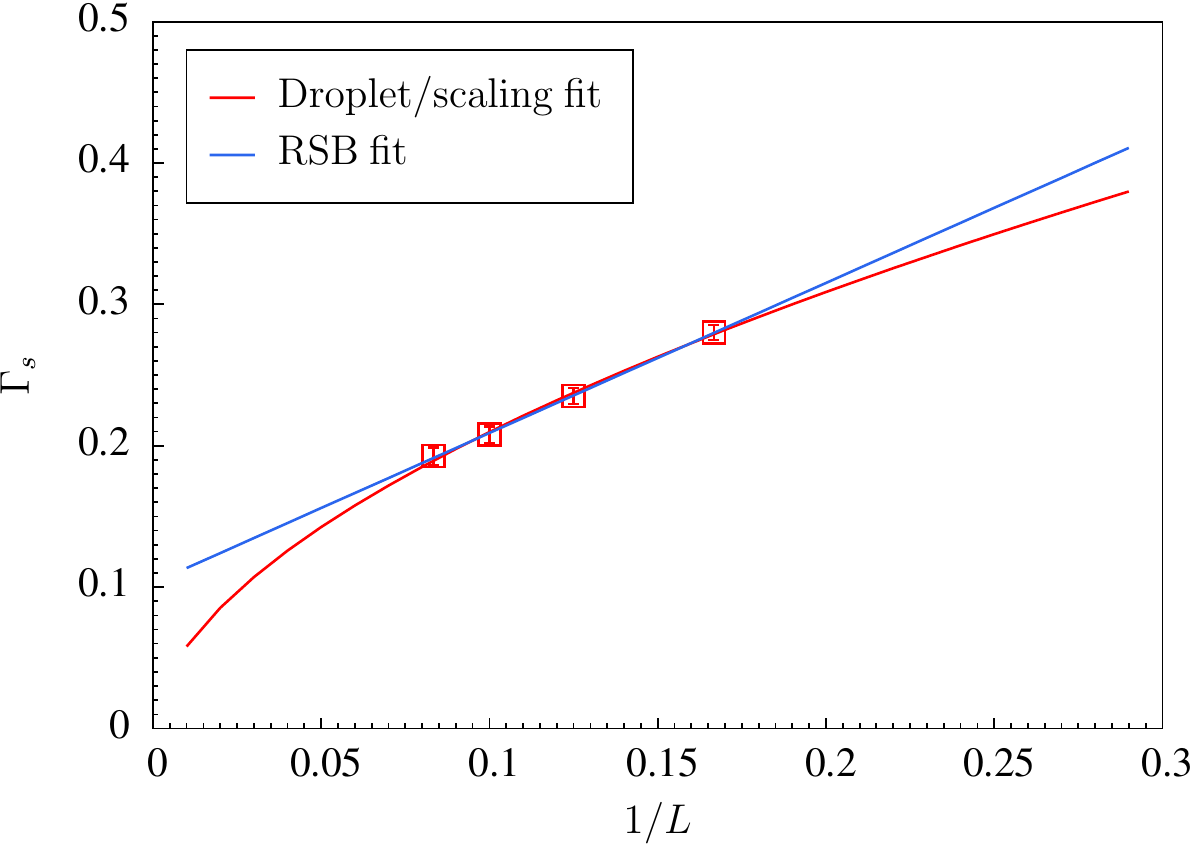}
\caption{
Comparison of the fits of $\GS$ assuming droplet/scaling (red curved
line, power law) or a RSB (blue straight line, constant and finite-size
correction) at $T=0.42$. Both functional forms fit the data comparably
well, i.e., $\GS$ does not distinguish the two pictures at these length
scales.  See text for details of fits.
}
\label{S1}
\end{center}
\end{figure}

The scaling of $\GS$ as a function of $L$ is shown in Fig.~\ref{Delta}.
A fit of the form $\GS = a (L/W)^{\ds-d}$ yields $\ds=2.44(3)$ and
$a=0.350(6)$ for $T=0.42$ and $\ds=2.54(3)$ and $a=0.378(6)$ for
$T=0.2$, with quality of fit \cite{press:95} $Q=0.74$ and $0.80$,
respectively.  Estimates of $\ds$ are in reasonable agreement with
previous results (e.g., Ref.~\cite{katzgraber:01}) before extrapolating
the aforementioned values to zero temperature. This suggest that the
relative domain wall might be deflecting out of the observation window
as $L$ becomes much larger than $\W$.  If the trend continues, this
would suggest a fractal domain wall and a single pair of pure states.
However, a fit with the assumption of space filling domain walls, i.e.,
$\ds=d$, of the form
\begin{equation}
\GS = a + b/(L/W) \to a \,\,\, {\rm for} \,\,\, L \rightarrow \infty
\label{eq:fitfill}
\end{equation}
with $a=0.103(3)$ and $b=0.265(5)$ is of similar quality with $Q=0.95$.
The results of both fits at $T=0.42$ are shown in Fig.~\ref{S1}.
Therefore, the finite-size scaling of $\GS$ is not able to distinguish
between space-filling and non-space-filling domain walls at these length
scales.

\subsection{Link overlap}
\label{linkoverlap}

\begin{figure}[htb]
\begin{center}
\includegraphics[width=\columnwidth]{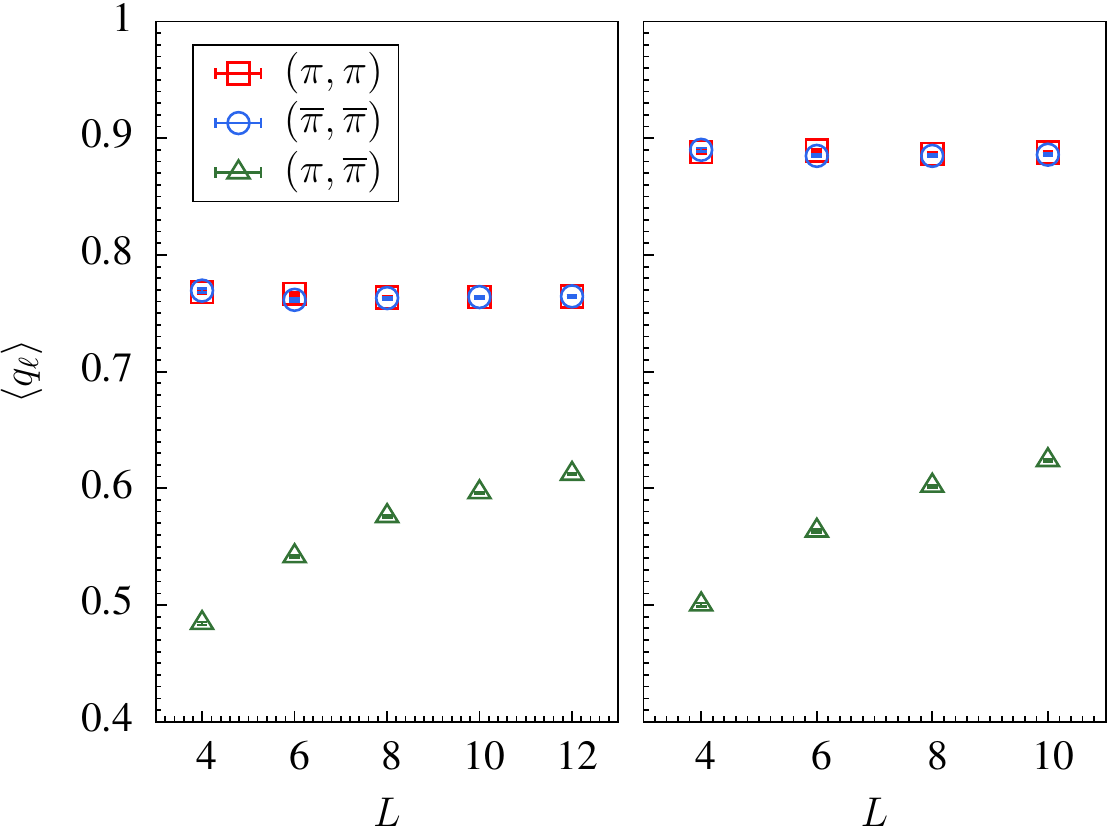}
\caption{
Average of the link overlap $\lav q_\ell \rav$ as a function of system
size $L$ for the three pairs of boundary conditions and temperatures
$T=0.42$ (left panel) and $T=0.2$ (right panel). Note that as $L$
increases the off-diagonal link overlap increases toward the diagonal
link overlap.
}
\label{ql}
\end{center}
\end{figure}

The average link overlap for diagonal and off-diagonal pairs of boundary
conditions for the full system are shown in Fig.~\ref{ql}. In agreement
with many previous studies, e.g., Ref.~\cite{katzgraber:01}, the average
diagonal link overlap is nearly independent of system size. On the other
hand, the average off-diagonal link overlap is an increasing function of
system size. If there is a single thermodynamic state and the relative
domain wall induced by the change of boundary conditions is not space
filling, then the diagonal and off-diagonal link overlap distribution
should become identical in the large volume limit. To test this
hypothesis, we consider the difference between diagonal and off-diagonal
link overlaps,
\begin{equation}
\GD=\llbracket \lav q_\ell 
\rav_{\pbc,\pbc} + 
\lav q_\ell \rav_{\abc,\abc}-\lav q_\ell \rav_{\pbc,\abc} -
\lav q_\ell \rav_{\abc,\pbc}\rrbracket /2 ,
\end{equation}
where the double brackets indicates  a disorder average.  The difference
between the diagonal and off-diagonal link overlaps is the average
volume occupied by the relative domain wall induced by changing boundary
conditions. Thus, if this domain wall has fractal dimension $\ds$ and we
compute $\GD$ in a system of size $L$, we again expect by box counting
\begin{equation}
\GD \sim L^{\ds-d} .  
\label{eq:dscale}
\end{equation}
Figure \ref{Delta} shows a log-log plot of $\GD$ as a function of $L$
and, from a power law fit, we estimate $\ds=2.43(1)$ at $T=0.42$ and
$\ds=2.57(2)$ at $T=0.2$ with quality of fit of $Q=0.098$ and $0.017$,
respectively. These values (including the temperature dependence) are
in agreement with the results of Ref.~\cite{wang:15a}. Furthermore, the
results at the lower temperature are in reasonable agreement with the
zero-temperature estimates of Refs.~\cite{palassini:00} and
\cite{katzgraber:01}.

\begin{figure}[htb]
\begin{center}
\includegraphics[width=\columnwidth]{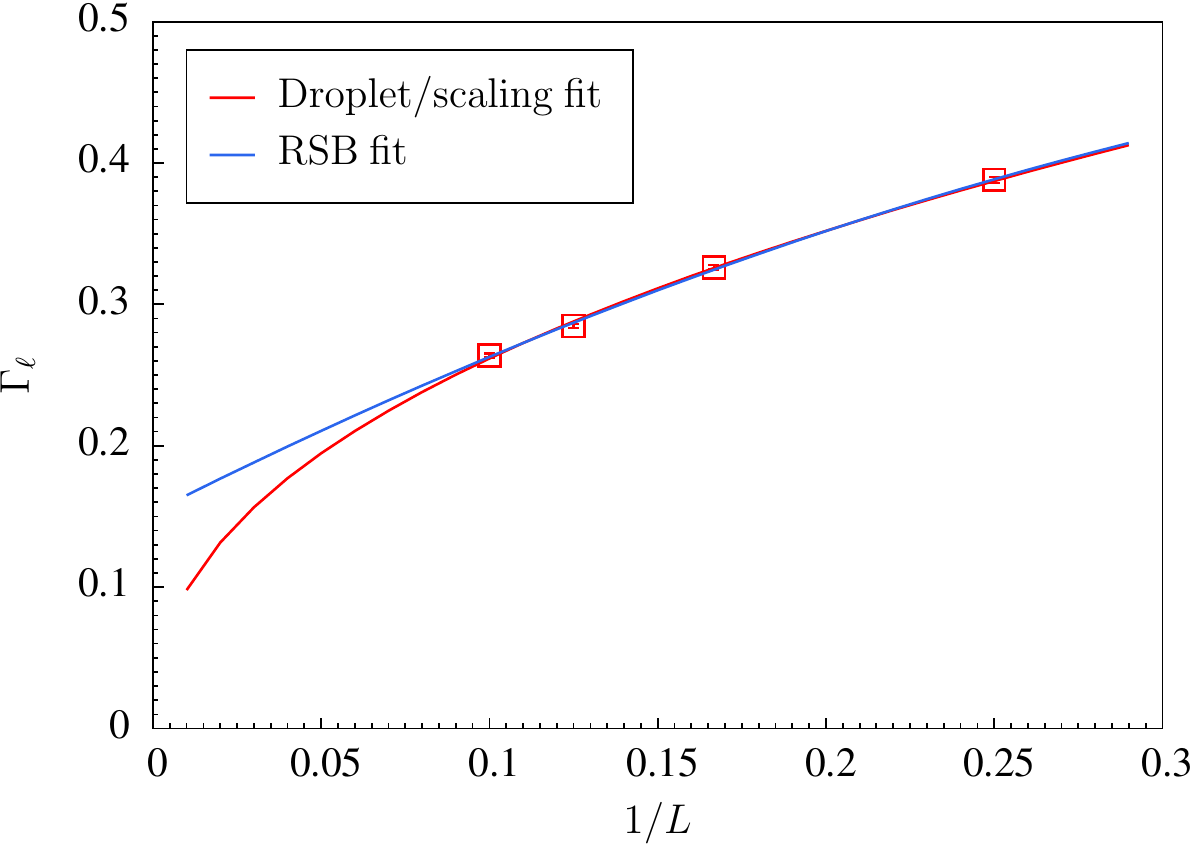}
\caption{
Comparison of the fits of $\GD$ assuming droplet/scaling (red curve,
power law) or a RSB (blue darker curve, constant and finite-size
corrections) at $T=0.2$.  Both functional forms fit the data comparably
well, i.e., $\GD$ does not distinguish the two pictures at these length
scales. See text for details of fits.
}
\label{S2}
\end{center}
\end{figure}

However, fits with $\GD \rightarrow a > 0$, implying space filling
domain walls (i.e., $\ds=d$) and multiple thermodynamic states are
similarly satisfactory. For $T=0.2$, a fit of the form $\GD= a + b/L +
c/L^2$ yields $a=0.15(2)$, $b=1.2(3)$ and $c=1.0(7)$ with $Q=0.03$.  The
uncertainty in $c$ is very large because of the shape of the error
ellipse of the three-parameter fit, but it should be noted that a
two-parameter fit with $c$ set to zero is of much lower quality although
it has a similar value of $a$. This RSB fit is shown in Fig.~\ref{S2}
along with the two-parameter power law fit $\GD = a L^{(\ds-3)}$ with
$a=0.7(2)$ and $\ds=2.57(2)$ with $Q=0.017$. Neither fit is high quality
and $\GD$ cannot distinguish the two scenarios at these length scales.

\section{Summary and future challenges}
\label{summary}

We have investigated the spin and link overlaps between the same and
different boundary conditions for the full system as well as in a
smaller observation window in the three-dimensional Edwards-Anderson
Ising spin glass. We find that the {\em off-diagonal} spin overlap
function for the full system is relatively flat and nearly independent
of system size, in agreement with a single thermodynamic state
consisting of a pair of pure states. This metric not used to date
represents another way to differentiate different theoretical scenarios.

On the other hand, as has been noted in previous studies, the {\em
diagonal} spin overlap function is nonzero and nearly independent of
systems size near $q=0$, which supports scenarios with many
thermodynamic states. The two sets of results together imply that for
system sizes currently accessible to low-temperature simulations, the
average spin overlap is incapable of distinguishing the two scenarios
for the three-dimensional Edwards-Anderson model.  A detailed analysis
of the link overlap and the spin overlap in a window is consistent with
fractal domain walls that have a fractal dimension near $\ds \approx
2.5$. However, we cannot rule out space filling domain walls.

While the use of $1/L$ finite-size corrections is reasonable for the
studied system sizes and hence we cannot rule out space-filling domain
walls, we believe it is an important future challenge to find a
theoretical basis for the used scaling form.  In contrast, the
ground-state energy per spin $e$ has corrections that scale as  $e = a +
b/L^x$, where $x=d-\theta \approx 2.76$ in three dimensions
\cite{boettcher:12}.  This exponent is much larger, and using such a
large exponent, the space-filling fits for our data are no longer
satisfactory. While it is possible that different quantities may show
different finite-size corrections at the zero-temperature fixed point,
it is still important to find a theoretical basis to explain why the
$1/L$ corrections are needed when studying the scaling of the link
overlap, yet not for the ground-state energy per spin.

Substantially larger system sizes would be required to clearly determine
whether domain walls induced by changing boundary conditions are space
filling or not using average spin overlaps in windows or link overlaps.
Since the first large-scale simulations in 2001 approximately 15 years
have passed. Betting on Moore's Law \cite{moore:65} we expected to be
able to revisit this problem and bring more clarity into the different
theoretical descriptions of the spin-glass state. However, our results
clearly show that more effort needs to be put into the development of
better algorithms, as well as new statistics to tackle these problems.

\begin{acknowledgments}

W.W., H.M.B and H.G.K. acknowledge support from the National Science
Foundation (Grant No.~DMR-1151387). J.M.~acknowledges support from the
National Science Foundation (Grant No.~DMR-1507506). We would like to
thank E.~Marinari, M.~A.~Moore, G.~Parisi, and D.~Stein for comments and
discussions.  H.G.K.~thanks Paul Hobbs for providing multiple sources of
inspiration.  The work of H.G.K., H.M.B.,~and W.W.~is supported in part
by the Office of the Director of National Intelligence (ODNI),
Intelligence Advanced Research Projects Activity (IARPA), via MIT
Lincoln Laboratory Air Force Contract No.~FA8721-05-C-0002. The views
and conclusions contained herein are those of the authors and should not
be interpreted as necessarily representing the official policies or
endorsements, either expressed or implied, of ODNI, IARPA, or the
U.S.~Government. The U.S.~Government is authorized to reproduce and
distribute reprints for Governmental purpose notwithstanding any
copyright annotation thereon.  We thank Texas A\&M University for access
to their Ada and Curie clusters.

\end{acknowledgments}

\bibliographystyle{apsrevtitle}
\bibliography{refs}

\end{document}